\documentclass[manuscript,screen]{acmart}
\usepackage{enumitem}
\usepackage{subcaption}
\usepackage{arydshln}
\usepackage{array}
\usepackage{algpseudocode}
\usepackage{amsmath}
\usepackage{graphicx}
\usepackage{float}
\usepackage{booktabs}
\usepackage{diagbox}
\usepackage{makecell}
\usepackage{bbding}
\usepackage{multirow}
\usepackage{multicol}
\usepackage{xspace}
\usepackage{color, soul}
\usepackage{makecell}

\usepackage{geometry}
\newcommand{\sysname}{\textrm{\textit{ProCLIP}}\xspace}

\AtBeginDocument{%
  }
\begin{document}

\title{Prompt-aware of Frame Sampling for Efficient Text-Video Retrieval}


\author{Deyu Zhang}
\email{zdy876@csu.edu.cn}
\orcid{0000-0003-0298-7868}
\affiliation{
  \institution{School of Computer Science and Engineering, Central South University}
  \city{Changsha}
  \state{Hunan}
  \country{China}
}

\author{Tingting Long}
\affiliation{
  \institution{School of Computer Science and Engineering, Central South University}
  \city{Changsha}
  \state{Hunan}
  \country{China}
}
\email{tingtinglong@csu.edu.cn}
\orcid{0009-0008-1745-3280}

\author{Jinrui Zhang}\authornote{Corresponding author}
\affiliation{
  \institution{Department of Computer Science and Technology, Tsinghua University}
  \state{Beijing}
  \country{China}}
\email{jinruizhang@tsinghua.edu.cn}
\orcid{0000-0001-9035-3050}

\author{Ligeng Chen}
\affiliation{
  \institution{Operation System Group, Honor Device Co., Ltd}
  \state{Beijing}
  \country{China}}
\email{chenlg@smail.nju.edu.cn}
\orcid{0000-0002-00076-3708}

\author{Ju Ren}
\affiliation{
  \institution{Department of Computer Science and Technology, Tsinghua University}
  \state{Beijing}
  \country{China}}
\email{renju@tsinghua.edu.cn}
\orcid{0000-0003-2782-183X}

\author{Yaoxue Zhang}
\affiliation{
  \institution{Department of Computer Science and Technology, Tsinghua University}
  \state{Beijing}
  \country{China}}
\email{zyx@tsinghua.edu.cn}
\orcid{0000-0001-6717-461X}

\renewcommand{\shortauthors}{Deyu Zhang et al.}

\begin{abstract}
Enabling efficient text-video retrieval on edge-end devices is critical for real-world applications. Yet, existing methods face a critical challenge in balancing accuracy and computational efficiency: uniform frame sampling methods ensure content coverage but incur prohibitive computational costs, while salient-frame sampling methods reduce overhead but suffer from query-agnostic frame selection that biases retrieval results. To address this, we propose \sysname, a user-centric framework that achieves state-of-the-art accuracy with significantly improved efficiency. We design a prompt-aware frame sampling strategy that dynamically guides lightweight feature extractors using textual prompts to select semantically relevant frames, overcoming the limitations of existing salient-frame sampling methods which rely on static, query-agnostic selection criteria. Moreover, we adopt a two-stage candidate pruning strategy that combines rapid coarse filtering via a lightweight module with CLIP-powered fine-grained re-ranking, enhancing retrieval efficiency while preserving accuracy. Experiments across benchmarks show \sysname achieves 75.3\% latency reduction versus baselines while maintaining competitive accuracy, i.e., R@1=49.0 in MSR-VTT dataset. \textit{Code is available at \href{https://github.com/tiffylong/ProCLIP}{\texttt{here}}.}

\end{abstract}





\maketitle

\section{Introduction} 
\label{introduction}
With the explosive growth of short-video platforms (e.g., TikTok) and user-generated content ecosystems\cite{zhang2024mobile,ghimire_survey_2022}, locally deployable text-to-video retrieval on edge-end devices has emerged as a critical enabler for real-time content creation\cite{yuan2025survey,wang2025cross,liang2024foundations,conde2024ais,wang2024end,qu2025mobile,lin2022text}. Specifically, video creators require precise semantic-level retrieval (e.g., searching for ``sunset beach surfing'' clips) directly on edge devices from massive raw footage (e.g., long videos), followed by instant preliminary editing and refinement\cite{nguyen2024videoclip,shen2025spatio,han2023bic,han2022adversarial,willis2025effects,Zhang2025MultimodalSemantic}. This paradigm addresses privacy-sensitive concerns inherent in cloud-based retrieval while ensuring seamless creative workflows in low-bandwidth or offline scenarios.

\begin{figure}[t]
    \centering
    \includegraphics[scale=0.45]{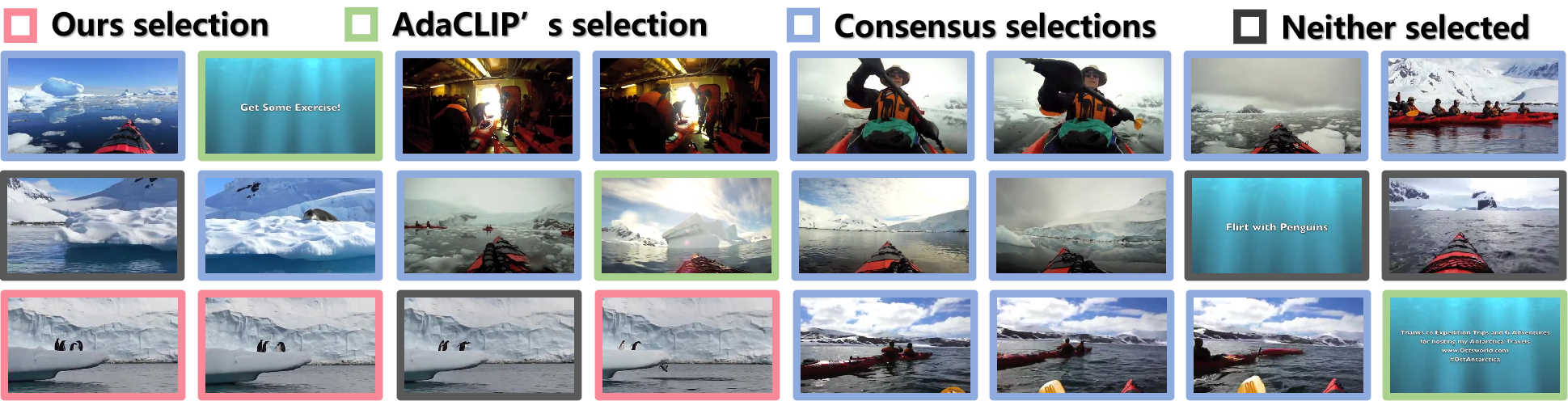}
    \caption{Frame selection comparison on an example video from the ActivityNet dataset~\cite{caba2015activitynet} for the user-provided prompt: ``A group kayaks through maze-like, ice-dotted waterways flanked by snow-capped rocks, encountering a seal perched on an iceberg. The kayakers also pass a group of penguins jumping into the water.'' Compared to AdaCLIP~\cite{hu2023adaclip}, which relies on static saliency criteria for frame selection, \sysname employs a prompt-aware sampling strategy that dynamically highlights frames aligned with the specific semantic cues in the user query—such as the seal and penguins—thereby capturing fine-grained visual concepts more effectively and producing frame selections that more faithfully reflect user intent and enhance retrieval accuracy. }
    \Description{xxxx}
    \label{fig:teaser}
\end{figure}

Current mainstream text-video retrieval methods primarily employ large-scale pre-trained models (e.g., CLIP\cite{radford2021learning}) to extract features from texts and videos, projecting them into a unified vector space for similarity-based retrieval\cite{luo2022clip4clip,fang2021clip2video,yu2025she,yu2024tf}. Based on video sampling strategies, existing approaches mainly fall into two categories: (1) Frame-by-frame or uniform video sampling. These methods\cite{ma2022x,gorti2022x,wang2023unified} typically sample numerous frames to ensure broad content coverage. Yet, \textbf{these methods incur substantial computational costs, particularly for long videos}. Benchmark tests on X-Pool\cite{gorti2022x} under an NVIDIA RTX 3090 GPU demonstrate that processing 1,000 video clips (30s duration each) incurs end-to-end inference latency exceeding 8s for a single retrieval operation. When handling longer-form videos (>100s duration) of equivalent dataset size, inference latency increases substantially to over 30s. Such delays severely degrade user experience in real-time interactive scenarios. 
(2) Salient frame-based sampling. To reduce computational costs, researchers propose efficient retrieval methods using salient frames\cite{hu2022mmsampler,hu2023adaclip}, leveraging lightweight visual feature extractors to assess frame ``saliency'' or ``information content''. These methods extract features from selected salient frames while jointly optimizing frame selection and retrieval tasks end-to-end. However, \textbf{a key limitation is that the criteria for selecting salient frames are unrelated to the user’s query content}. This mismatch between selected ``informative'' frames and user-relevant content is evident in Figure \ref{fig:teaser}: for multi-scene videos (e.g., containing both ``kayaking person'' and ``penguin behaviors''), existing methods tend to prioritize identical frame sequences (e.g., ``person''-related frames) across queries, leading to insufficient information about non-dominant scenes (e.g., ``penguins'') and biased similarity computation that causes retrieval failures.

\begin{figure}[t]
\centering
    \includegraphics[width=0.6\textwidth]{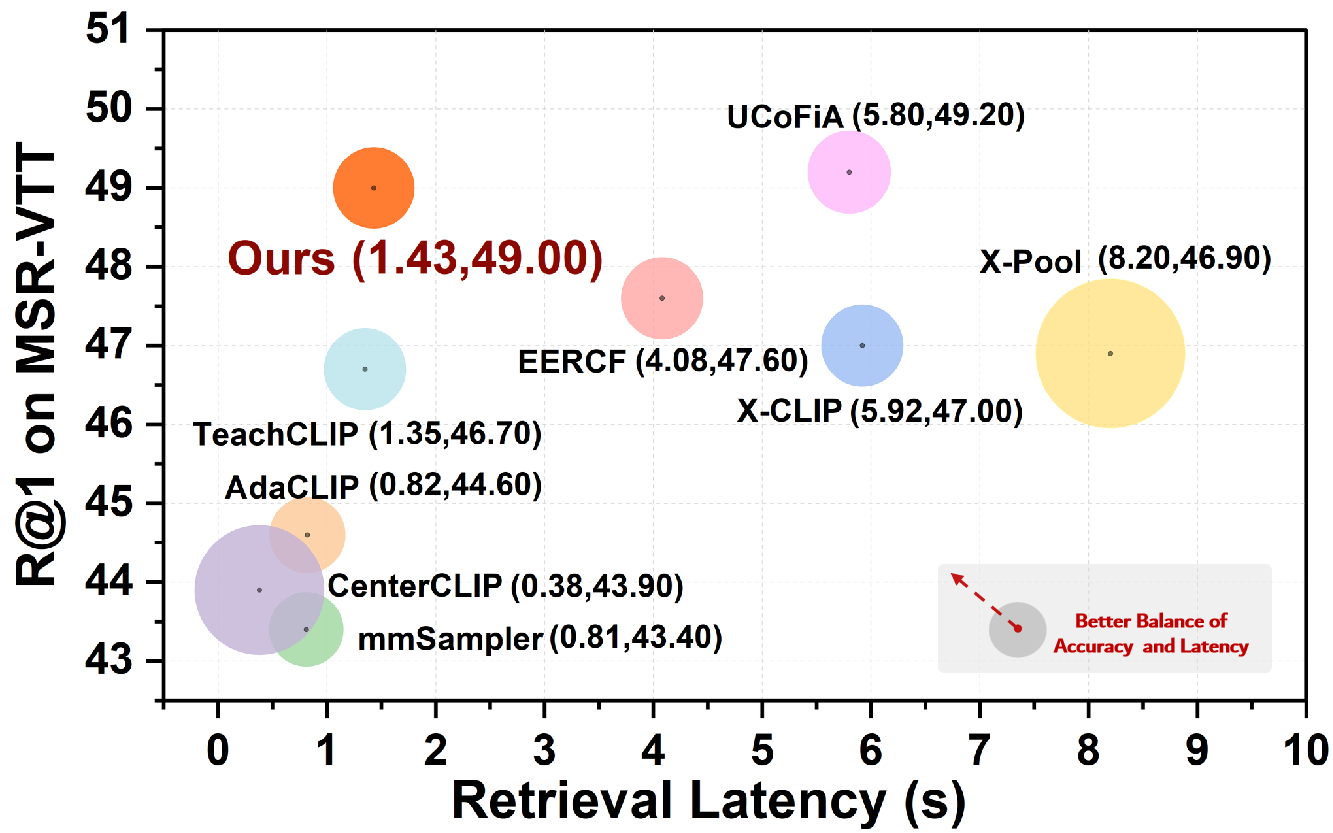}
    \caption{Retrieval Latency-Accuracy Trade-off on MSR-VTT test set. X/Y-axes: retrieval latency (seconds) vs. R@1. Each method is shown as a bubble, where its size reflects model's size. Latency is measured on a single RTX 3090 GPU. Compared with baselines, our method (computation ratio $k = 50\%$, see in Sec. \ref{3-2}) achieves Pareto-optimal performance, i.e., moving toward upper-left indicates better performance.}
    \Description{xxxx}
    \label{efficiency1}
\end{figure}

To address the challenge of balancing accuracy and efficiency, we propose \sysname, a user-centric framework for text-video retrieval. \sysname achieves comparable accuracy to SoTA methods while significantly reducing computational overhead. We design a prompt-aware frame sampling strategy, which integrates user prompts to dynamically guide lightweight visual feature extractors in frame selection, overcoming the limitations of fixed visual saliency criteria in existing approaches. Additionally, we design a two-stage candidate pruning strategy to mitigate the computational cost of dynamic evaluation: first, lightweight feature extractors rapidly filter out irrelevant videos, and then CLIP is applied for fine-grained ranking on the high-confidence candidate set.

However, the core challenge in designing \sysname lies in \textbf{the insufficient cross-modal alignment capability of lightweight visual feature extractors}. Traditional lightweight extractors (e.g., MobileNet-V1/V2/V3\cite{howard2017mobilenets,sandler2018mobilenetv2,howard2019searching}) are optimized for single-modal tasks (e.g., ImageNet classification\cite{deng2009imagenet}) and lack the large-scale cross-modal contrastive learning priors inherent to CLIP. This limitation results in two key cross-modal misalignment issues: firstly, due to the absence of text-visual training, lightweight extractors struggle to establish semantic associations between user queries and visual content, rendering frame selection unable to accurately reflect user intent; secondly, during the initial screening stage, the visual features extracted by lightweight models are misaligned with the text features generated by CLIP, as they lack a shared semantic embedding space. This misalignment leads to significant discrepancies between the ``high-confidence candidate set'' and the true matching set, where noise in the candidate set negates the acceleration benefits of the screening stage and compromises retrieval accuracy.

To address the above issues, our prompt-aware frame sampling design decouples the word-level and sentence-level semantics of the user prompt, calculating attention scores with video frames separately. We design a dynamic gating strategy to achieve multi-granularity fusion. Through end-to-end joint training with the retrieval task, this strategy enhances the lightweight frame feature extractor's ability to align accurately with user query semantics. For feature space misalignment in two-stage candidate pruning, we propose a cross-modal feature distillation technique using $MSE$ loss. Building upon the lightweight visual feature extractor, we design a compact Transformer encoder that minimizes $L2$ distance between its outputs and CLIP visual features, enabling implicit inheritance of CLIP's cross-modal alignment capability. This dual approach improves candidate screening quality while maintaining computational efficiency.

Extensive experiments on mainstream video retrieval datasets demonstrate that \sysname achieves an excellent trade-off between efficiency and accuracy. As shown in Figure~\ref{efficiency1}, our approach maintains \textbf{SoTA-level R@1 performance} while reducing retrieval latency by \textbf{75.3\%}. Compared with existing efficient retrieval methods, \sysname improves accuracy up to \textbf{12.9\%} while maintaining comparable latency, showing significant performance advantages.

In summary, the main contributions are as follows:
\begin{itemize}[leftmargin=0pt, itemindent=*, nosep]
\item {We propose \sysname, an efficient and accurate text-to-video retrieval framework optimized for user-side edge devices, delivering responsive and resource-friendly search experiences.}

\item{We propose prompt-aware frame sampling strategy, which selects the frames most relevant to the user query, enhancing text-to-video retrieval by addressing the limitations of static visual saliency criteria in existing methods.}

\item{We design two-stage candidate pruning strategy that integrates a CLIP-distilled lightweight module for coarse filtering with CLIP-based fine-grained re-ranking, significantly improving retrieval efficiency without sacrificing accuracy.}

\item{We demonstrate \sysname achieves significant improvement in both retrieval speed and accuracy, achieving a 75.3\% reduction in retrieval latency while boosting accuracy up to 12.9\% compared to baselines, highlighting the superior balance between performance and efficiency.}

\end{itemize}

\section{Related Works} 
\textbf{Uniform sampling-based Text-Video Retrieval.} Text-video retrieval aims to retrieve relevant videos given textual queries, where uniform frame sampling serves as the foundational paradigm for video representation. In this framework, videos are first decomposed into fixed-interval frames through uniform temporal sampling, then aligned with textual features in a shared embedding space. The dominant approach encodes uniformly sampled video frames and text into feature vectors, where similarity computation enables cross-modal matching. 
Existing methods\cite{gorti2022x,ma2022x,wang2023unified} process uniformly sampled frames through CLIP's image encoder, then aggregate frame features via temporal pooling or attention strategies. This inherits CLIP's alignment capability while maintaining computational efficiency through deterministic frame sampling. Recent advancements refine alignment strategies within this framework. X-CLIP\cite{ma2022x} implements cross-grained contrast through frame-word interactions on uniformly sampled visual features, while UCoFiA\cite{wang2023unified} improves retrieval accuracy through a more refined coarse-to-fine hierarchical alignment strategy. Despite their decent accuracy, these methods suffer from high computational costs due to dense frame sampling.

\textbf{Salient frame sampling-based Text-Video Retrieval.} 
To alleviate the computational burden of video feature extraction, CenterCLIP\cite{zhao2022centerclip} propose a token clustering strategy to eliminate redundancy by merging homogeneous tokens from adjacent frames, reducing computational overhead. mmSampler\cite{hu2022mmsampler} and AdaCLIP\cite{hu2023adaclip} propose salient-frame-based solutions. These methods employ lightweight CNNs to assess the ``information content'' or ``saliency'' of frames, selectively extracting those more informative frames (i.e., salient frames) for feature computation. In addition, they integrate frame selection and retrieval into an end-to-end optimization framework. However, since their saliency criteria are entirely learned from statistical patterns in training data, the selection strategy becomes static after model training, consequently failing to adapt dynamically to users' diverse semantic intents and query requirements.

\section{System Design}

\begin{figure}[htp]
    \centering
    \includegraphics[scale=0.22]{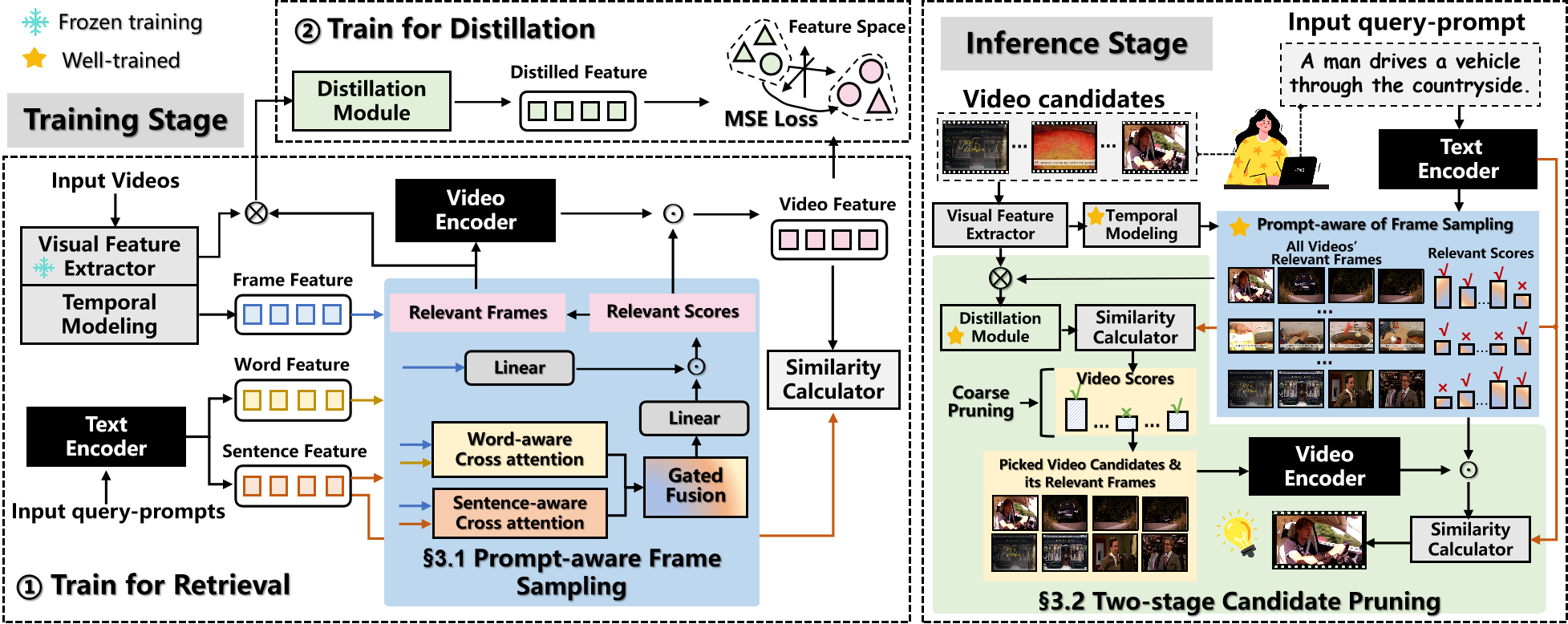}
    \caption{Training and inference pipelines of \sysname.}
    \label{framework}
\end{figure}

Figure \ref{framework} illustrates the training and inference pipeline of \sysname. In this section, we first outline the preliminary feature extraction process (Sec.~\ref{3-1}). Then, we design the prompt-aware frame sampling method, which enables the query prompt to guide relevant frame selection (Sec.~\ref{3-2}). We also present the two-stage pruning strategy to mitigate the online computational overhead (Sec.~\ref{3-3}). Finally, we describe \sysname's detailed training protocols and inference acceleration procedures (Sec.~\ref{3-4}).

\vspace{-8pt}
\subsection{Feature Extraction}
\label{3-1}
This section describes the feature extraction for both visual and textual modalities, providing essential support for subsequent prompt-aware frame sampling and two-stage candidate pruning.

\textbf{Visual Feature Extractor.} Given an input video comprising $\mathcal{N}$ frames, we first apply a lightweight visual feature extractor (e.g., MobileNet-V3) to extract frame-level feature representations. These features undergo dimensionality reduction through a learnable linear projection layer, mapping them into a latent subspace. Subsequent to the injection of sinusoidal positional embeddings, the encoded frame representations are propagated through a multi-layer temporal dependency modeling module. This module employs a depth-variable Transformer architecture: a 3-layer self-attention network is instantiated for short-duration videos ($\mathcal{T}\leq 60$ seconds), while a deeper 5-layer configuration is activated for long-form content ($\mathcal{T}> 60$ seconds) to enhance multi-scale temporal context aggregation. The resultant frame-wise embeddings serve as the basis for downstream frame selection operations.

\textbf{Text Encoder.} Given a user query prompt \(Q\), we leverage CLIP's text encoder to extract token-wise feature representations. Specially, the encoder outputs a sequence of word embeddings  \(\mathbf{W}=[W_1,W_2,...,W_W] \in \mathbb{R}^{W \times D}\), where \(W\) represents the token sequence length and \(D\) denotes the embedding dimension. Following standard practice in  \cite{wang2023unified,radford2021learning}, we take the representation of the [EOS] token embedding as the sentence representation, denoted as \(\mathbf{S} \in \mathbb{R}^{1 \times D}\), where \(D\) represents the sentence dimension.

\textbf{Video Encoder.} Given a video sequence containing \(N\) frames, we derive \(K\) query-aware salient frames $ \left\{ F_k\right\}^{K}_{k=1}$ with associated saliency weights $ \left\{ \alpha _k\right\}^{K}_{k=1}$ through the proposed prompt-aware frame sampling module.  The compressed video representation is thus formulated as \(V = [F_1, F_2, ..., F_K] \in \mathbb{R}^{K \times D_v}\), where $D_v$ denotes the visual feature dimension. Following  recent vision-language alignment approaches \cite{hu2023adaclip,wang2023unified}, we employ CLIP's visual encoder to extract frame features $E_v \in \mathbb{R}^{K \times D}$, then modulate these features by their corresponding $\alpha_k$ values through element-wise multiplication. The weighted features are subsequently aggregated via a lightweight Transformer layer (single-head self-attention with LayerNorm) to produce the final video embedding $v \in \mathbb{R}^D$.

\subsection{Prompt-aware Frame Sampling}
\label{3-2}

To accurately guide relevant frame selection using the user prompt, we design a prompt-aware frame sampling strategy that incorporates both word-aware and sentence-aware cross-attention strategy. Specifically, the word-aware cross-attention captures fine-grained semantic interactions between individual textual tokens and video frame features through token-level similarity computation, while the sentence-aware cross-attention extracts high-level contextual relevance between the aggregated sentence embedding and global video context. The outputs of these two attention strategies are adaptively fused using a learnable gating function that optimizes their relative contributions during training. This produces unified frame relevance scores that jointly consider local lexical matching and global semantic coherence, thereby ensuring the selected frames  $ \left\{ F_k\right\}^{K}_{k=1}$ optimally satisfy the user's query intent  \(Q\).

\begin{figure}[htbp]
    \centering
    \subfloat[Word-aware cross-attention]{\includegraphics[width=0.45\linewidth]{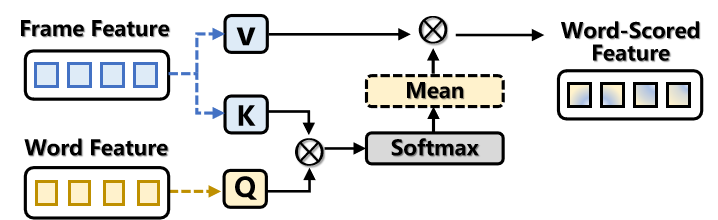}}
    \hspace{0.5em} 
    \subfloat[Sentence-aware cross-attention]{\includegraphics[width=0.45\linewidth]{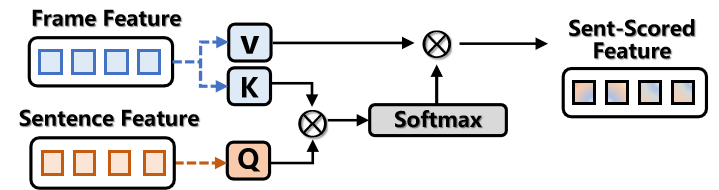}}
    \caption{Word-aware cross attention and sentence-aware cross attention of prompt-aware frame sampling.}
    \label{word_and_sent}
\end{figure}

\textbf{Word-aware Cross Attention.}  We design word-aware cross attention to capture the fine-grained interactions between textual features and video frames, as illustrated in Figure~\ref{word_and_sent} (a). Given word features \(\mathbf{W} \in \mathbb{R}^{ W \times D}\) and frame features \(\mathbf{F} \in \mathbb{R}^{ N \times D}\), where \(N\) is the number of frames, we first compute the word-to-frame attention scores using scaled dot-product attention:  
\begin{equation}  
\setlength\abovedisplayskip{3pt}
\setlength\belowdisplayskip{3pt}
\mathbf{A} = \text{softmax} \left( \frac{\mathbf{W} \mathbf{F}^T}{\sqrt{D}} \right), \quad \mathbf{A} \in \mathbb{R}^{ W \times N}.  
\end{equation}  

Each entry \(A_{ij}\) represents the attention weight of the \(i\)-th word on the \(j\)-th frame, indicating how much a specific word attends to a particular frame. To aggregate relevance across all words, we compute the mean attention score across the word dimension:  
\begin{equation}  
\setlength\abovedisplayskip{3pt}
\setlength\belowdisplayskip{3pt}
\mathbf{S} = \frac{1}{W} \sum_{i=1}^{W} A_{ij}, \quad \mathbf{S} \in \mathbb{R}^{N}.  
\end{equation}  

These word relevance scores \(\mathbf{S}\) quantify the word-level importance of each frame. Finally, we use them to reweight the original frame features, ensuring that frames more relevant to the textual query receive higher contributions:  
\begin{equation} 
\setlength\abovedisplayskip{3pt}
\setlength\belowdisplayskip{3pt}
\mathbf{W_o} = \mathbf{F} \cdot \mathbf{S}^T, \quad \mathbf{W_o} \in \mathbb{R}^{ N \times D}.  
\end{equation}  

Through this strategy, word-aware cross attention dynamically assigns higher importance to frames that contain more semantically relevant content based on the textual query. This selective reweighting enhances retrieval precision by emphasizing frames that align closely with fine-grained textual cues while suppressing less informative or irrelevant frames.

\textbf{Sentence-aware Cross Attention.}  Sentence-aware cross attention holistically assesses the relevance of each frame to the user query. Given a sentence feature \(\mathbf{S} \in \mathbb{R}^{1 \times D}\) and frame features \(\mathbf{F} \in \mathbb{R}^{N \times D}\), as shown in Figure~\ref{word_and_sent} (b), the process is similar to the word-aware cross attention but operates at the sentence level.  First, we compute the sentence-to-frame attention scores using scaled dot-product attention, and then normalize these scores along the frame dimension using the softmax function to obtain sentence relevance scores. These scores reflect how well each frame aligns with the user query at the sentence level. The resulting relevance scores are then used to refine the frame features, enhancing the frame selection process by emphasizing frames that are more semantically relevant to the sentence:
\begin{equation}  
\setlength\abovedisplayskip{3pt}
\setlength\belowdisplayskip{3pt}
\mathbf{S_o} = \mathbf{F} \cdot \text{softmax} \left( \frac{\mathbf{S} \mathbf{F}^T}{\sqrt{D}} \right)^T, \quad \mathbf{S_o} \in \mathbb{R}^{ N \times D}.  
\end{equation}

\textbf{Gated Attention Fusion.} 
To effectively integrate information from word-level and sentence-level attention outputs, we employ a gated attention fusion. Given the word-aware cross-attention output $\mathbf{W_o} \in \mathbb{R}^{N \times D}$ and the sentence-aware cross-attention output $\mathbf{S_o} \in \mathbb{R}^{N \times D}$, we first concatenate them to form a combined feature representation:  
\begin{equation}  
\setlength\abovedisplayskip{3pt}
\setlength\belowdisplayskip{3pt} 
    \mathbf{x} = [\mathbf{W_o}; \mathbf{S_o}], \quad \mathbf{x}\in \mathbb{R}^{N \times 2D}.
\end{equation}  

To adaptively determine the contribution of each feature, we design a gating function $g(\cdot)$, parameterized by a two-layer feedforward network with a ReLU activation and a sigmoid function:  
\begin{equation}
\setlength\abovedisplayskip{3pt}
\setlength\belowdisplayskip{3pt} 
    g = \sigma\left( \text{ReLU}(\mathbf{x} \mathbf{W}_1^\top + \mathbf{b}_1) \mathbf{W}_2^\top + b_2 \right), \quad g \in \mathbb{R}^{N \times 1}.
\end{equation}  
where $\mathbf{W}_1 \in \mathbb{R}^{D \times 2D}$ and $\mathbf{b}_1 \in \mathbb{R}^{D}$ are the weight and bias of the first linear layer, and $\mathbf{W}_2 \in \mathbb{R}^{1 \times D}$ and $b_2 \in \mathbb{R}$ are those of the second layer. Note that the weights $\mathbf{W}_1$ and $\mathbf{W}_2$ are shared across all $N$ frames, and the batch dimension $N$ is preserved through matrix multiplication. The sigmoid function $\sigma(\cdot)$ ensures that the gating value $g$ lies within the range $(0,1)$, enabling a learnable balance between word-level and sentence-level features.

The final fused representation $\mathbf{y}$ is a weighted sum of the two attention outputs, which enables the model to dynamically adjust the relative importance of word-level and sentence-level information based on the input context.  
\begin{equation}  
\setlength\abovedisplayskip{3pt}
\setlength\belowdisplayskip{3pt}
    \mathbf{y} = g \cdot \mathbf{W_o} + (1 - g) \cdot \mathbf{S_o}, \quad \mathbf{y} \in \mathbb{R}^{N \times D}.
\end{equation}  

After obtaining the fused representation $\mathbf{y}$, we use it to guide the selection of relevant frames. Following \cite{hu2023adaclip}, a multi-layer perceptron (MLP) maps the initial frame features and $\mathbf{y}$ to their respective logits and relevance scores, whose element-wise product yields the final frame scores. During training, we adopt the differentiable Hard Top-K algorithm from \cite{hu2023adaclip} to enable end-to-end optimization of the frame selection module along with the retrieval task.

\subsection{Two-stage Candidate Pruning}
\label{3-3}
Utilizing user prompts to guide the selection of relevant frames overcomes the limitations of fixed saliency criteria in existing methods. Yet, it requires recomputing relevant frames for every query, leading additional online computational overhead. To address this, we propose two-stage candidate pruning strategy: Firstly, we employ lightweight feature extractor to prune irrelevant videos rapidly. Then, we exclusively apply CLIP to the high-confidence candidate set for fine-grained ranking. This strategy significantly reduces the number of videos requiring computation, thereby lowering the overall computational cost without sacrificing retrieval accuracy.

\textbf{Initial Pruning with Lightweight Feature Extractor.} In the initial pruning stage, we compute the similarity scores $R(v_i, q) = \text{sim}(\mathbf{F}_i, \mathbf{Q}_q)$ between the CLIP-generated sentence features $\mathbf{S}_q$ (for query $q$) and all video frame features $\mathbf{F}_i$ (for video $v_i$). We dynamically retain the top $k\%$ of videos ranked by similarity scores. The candidate set $S_{\text{init}}$ is constructed as:
\begin{equation}
\setlength\abovedisplayskip{3pt}
\setlength\belowdisplayskip{3pt}
    S_{\text{init}} = \left\{ v_i \mid R(v_i, q) \text{ ranks in top } k\% \text{ of all videos} \right\},
\end{equation}
where $k$ is a tunable hyperparameter controlling the pruning intensity. During inference, $k$ can be adjusted to explore the trade-off between retrieval precision (higher $k$ preserves more candidates) and computational efficiency (lower $k$ reduces downstream processing).


\textbf{Feature Distillation.} A fundamental challenge in the initial pruning stage stems from the modality gap between lightweight backbone networks and CLIP's cross-modal alignment capacity.  The high-recall candidate pool generated by the lightweight extractor exhibits substantial distributional shift from CLIP's semantic-aware feature space, potentially leading to a high rate of false positives and false negatives during coarse-grained retrieval. 

To address this issue, we design a cross-modal knowledge distillation framework that aligns the lightweight network's representations with CLIP's video embedding space. This improves the quality of the candidate set by enhancing its alignment with CLIP’s learned feature space. Specifically, we incorporate a lightweight distillation module after the feature extractor, which consists of  a three-layer Transformer encoder stack  appended to the backbone network.   
During optimization, we freeze all pretrained parameters and exclusively optimize the distillation module to further enhance feature consistency. As shown in Fig. \ref{framework}, the distillation objective minimizes the discrepancy between student features (lightweight network) and teacher features (CLIP) through Mean Squared Error (MSE) loss ($\mathcal{L}_{\text{MSE}}$):
\begin{equation}
\setlength\abovedisplayskip{3pt}
\setlength\belowdisplayskip{3pt}
    \mathcal{L}_{\text{MSE}} = \frac{1}{N} \sum_{i=1}^{N} \left\| \phi(v_i) - \phi_{\text{CLIP}}(V_i) \right\|^2
\end{equation}
where $\phi(v_i)$ denotes the visual feature extracted by the lightweight distillation module, and $\phi_{\text{CLIP}}(V_i)$ represents CLIP’s final video feature. 

This representation alignment strategy mitigates approximation errors in the pruning stage, ensuring the candidate set maintains semantic congruence with CLIP's final retrieval space while preserving computational efficiency.

\subsection{Model Training and Inference}
\label{3-4}
This section delineates the training and inference pipeline of \sysname. The training comprises two stages: 1) joint end-to-end optimization of the prompt-aware frame sampling strategy and retrieval objective; 2) dedicated optimization of the distillation module within the two-stage candidate pruning strategy. During inference, the framework leverages hierarchical computational pruning to achieve efficiency gains while preserving accuracy. 

\textbf{Training Pipeline.} We divide the training process into two stages, as illustrated in Figure \ref{framework}. In the retrieval stage, we freeze the lightweight visual feature extractor and jointly optimize prompt-aware frame sampling and text-video retrieval in an end-to-end manner. Given a batch of $B$ video-text pairs during retrieval training, we extract video and text features and compute a similarity matrix of size $B \times B$ between all video-text pairs. We adopt symmetric cross-entropy loss ($\mathcal{L}$) to maximize the similarity between positive video-text pairs:
\begin{equation}
\setlength\abovedisplayskip{3pt}
\setlength\belowdisplayskip{3pt}
    \mathcal{L}_{v2t} = -\frac{1}{B} \sum_{i=1}^{B} \log \frac{\exp(s(\mathbf{v}_i, t_i))}{\sum_{j=1}^{B} \exp(s(\mathbf{v}_i, t_j))},
\end{equation}
\begin{equation}
    \mathcal{L}_{t2v} = -\frac{1}{B} \sum_{i=1}^{B} \log \frac{\exp(s(\mathbf{v}_i, t_i))}{\sum_{j=1}^{B} \exp(s(\mathbf{v}_j, t_i))},
\end{equation}
\begin{equation}
    \mathcal{L} = \frac{1}{2} (\mathcal{L}_{v2t} + \mathcal{L}_{t2v}),
\end{equation}
where $s(\mathbf{v}_i, t_j)$ represents the cosine similarity score between the $i$-th video and the $j$-th textual description within a batch of size $B$.

In the second stage, all pretrained parameters except the distillation module are frozen. Given a batch of $B$ video-text pairs,  we process CLIP-derived frame features alongside their lightweight counterparts through the distillation network, minimizing their L2-distance via MSE loss (defined in Sec.\ref{3-3}). This bridges the representational discrepancy between efficient and CLIP-based features.


\textbf{Inference Acceleration.} 
Given a user-provided query-prompt, multiple videos are first retrieved as potential matches. As shown in Figure \ref{framework}, the query and all video candidates undergo preliminary feature extraction, followed by prompt-aware frame sampling, where the query guides the selection of relevant frames and assigns frame Relevant Scores. This process discards redundant frames and reduces frame-level computation.

We then compute similarities between sentence-level text features and video features from selected frames, producing video scores for coarse pruning. Videos with low scores are filtered out, reducing the load of later CLIP-based processing.

Finally, \sysname processes a small set of top-ranked videos using CLIP's visual encoder. Aggregated video features are weighted by frame relevance before similarity computation, enhancing accuracy while preserving efficiency.

\section{Experiment}
\subsection{Experimental Detail}
\subsubsection{\textbf{Dataset}}
We evaluate \sysname on three public benchmarks: 

\textbf{MSR-VTT}\cite{xu2016msr} consists of 10,000 videos, each paired with 20 text captions. The duration of the videos varies between 10 and 32 seconds. Following \cite{luo2022clip4clip,hu2023adaclip}, we train \sysname on 9,000 videos and evaluate on a set of 1,000 selected video-text pairs.

\textbf{MSVD}\cite{chen2011collecting} consists of 1,970 videos, with durations ranging from 1 to 62 seconds. Each video is paired with approximately 40 captions. Following \cite{luo2022clip4clip}, we use 1,200 videos for training, 100 for validation, and 670 for testing. For evaluation, each video in the test set is matched with multiple text captions. 

\textbf{ActivityNet}\cite{caba2015activitynet} is a large-scale dataset comprising 20,000 YouTube videos and 100,000 captions, with an average video length of 180 seconds. Following \cite{luo2022clip4clip}, we concatenate all text captions of a video into a single paragraph and conduct paragraph-to-video retrieval. \sysname is trained on 10,000 videos and evaluated on the ‘val1’ split, which includes 5,000 videos.

All three public benchmarks comprise real-world, user-generated videos collected from open platforms such as YouTube, Tik Tok, capturing authentic everyday scenarios. To further validate \sysname's generalization capability in real-world applications, conduct two more personalized evaluations (see Sec.~\ref{sec:personalized evaluation}), including a query personalization test on ActivityNet and a user satisfaction study with personally captured videos (e.g., travel recordings, home videos).

\subsubsection{\textbf{Evaluation Metric}}
We assess retrieval performance using standard metrics: (1) \textbf{Recall@K(R@K)}. Proportion of queries where the ground-truth video appears in the top-K results, with higher values denoting superior accuracy. (2) \textbf{Mean Rank (MnR) }. Average ordinal position of correct matches in ranked lists, with lower values indicating tighter semantic alignment. 

In addition, we assess inference efficiency to evaluate the practical usability of \sysname in real-world scenarios (see Sec. \ref{sec:efficiency_performance}) with four metrics, including (1) \textbf{First Query Latency (FQ-Latency)}: The time (seconds) required to retrieve the Top-1 result for a single user query from 1,000 test videos, including the time for text feature encoding, video feature computation for all videos, and similarity calculation. The average latency over 1,000 queries is reported. (2) \textbf{Average Query Latency (AQ-Latency)}: The average time (seconds) over ten retrieval rounds, assuming that after the first retrieval, all video features are cached. This setting reflects a more realistic user experience, where multiple distinct queries are issued over the same video set. (3) \textbf{Average Query energy consumption (AQ-Energy)}: The average energy consumption (in joules) during AQ-Latency measurement, computed using NVIDIA's PyNVML toolkit. This metric reflects the power efficiency of the retrieval process and is critical for deployment on energy-constrained edge devices. (4) \textbf{Maximum GPU Memory Usage (Max GPU Mem)}: The peak GPU memory usage (in GB) during inference, measured using \texttt{$torch.cuda.max\_memory\_allocated()$}. This metric evaluates the memory footprint of each method and its suitability for deployment on resource-limited devices.

\subsubsection{\textbf{Baselines}}  
To provide a comprehensive evaluation with current text-to-video retrieval methods, we select three \textbf{uniform sampling-based methods}, including X-Pool (CVPR 2022)\cite{gorti2022x}, X-CLIP (MM 2022)\cite{ma2022x}, and UCoFiA (CVPR 2023)\cite{wang2023unified}, and five \textbf{latest efficient and salient-frame sampling methods}, e.g., mmSampler (MM 2022)\cite{hu2022mmsampler}, AdaCLIP (MM 2023)\cite{hu2023adaclip}, CenterCLIP (SIGIR 2022)\cite{zhao2022centerclip}, EERCF (AAAI 2024)\cite{tian2024towards} and TEACHCLIP (CVPR 2024)\cite{tian2024holistic}. These baselines are selected to cover both conventional sampling paradigms and recent advances in retrieval acceleration. Following the practice of prior works(e.g., EECRF and TeachCLIP), we strictly retrain and evaluate the baselines using their official code and reported training configurations. Training is terminated upon loss convergence to preserve benchmark accuracy. This is necessary as some pretrained models are not released, and several evaluation metrics are missing on certain datasets.

\subsubsection{\textbf{Implementation Details}} 
The model is initialized with the pre-trained weights of CLIP’s ViT-B/32. The lightweight visual feature extractor is based on MobileNetV3\cite{howard2019searching}, which is pretrained on ImageNet\cite{deng2009imagenet}. The distillation module employs a three-layer Transformer encoder with 8 attention heads and a feedforward dimension of 256, where input features are first projected to a unified dimension and enhanced with positional embeddings. The model is trained following AdaCLIP's HardTop-k algorithm\cite{hu2023adaclip}, with an initial temperature of 5 and an exponential decay factor of 0.045. We uniformly sample 16, 16, and 64 frames for MSR-VTT, MSVD, and ActivityNet, respectively, with maximum text token lengths of 32, 32, and 64. Training runs for 30 epochs using Adam, with batch sizes of 32, 32, and 16 per GPU and an initial learning rate of 1e-7 for the CLIP backbone and 1e-4 for other modules, decayed via a cosine schedule. Retrieval experiments are conducted on two NVIDIA A6000 GPUs, while efficiency evaluations run on a single NVIDIA RTX 3090 GPU. This hardware selection aligns with our deployment target of laptop and desktop-grade edge devices, where the RTX 3090 serves as a prevalent real-world baseline for mid-range GPU deployment in text-to-video retrieval systems.

\subsection{Results on Common Benchmarks}
\subsubsection{\textbf{Retrieval Performance on Common Benchmarks}}
Table~\ref{tab:Retrieval} presents a comprehensive comparison between \sysname and existing approaches on common benchmarks. The upper section of the table includes \textit{uniform sampling-based methods}, while the lower section focuses on \textit{efficient retrieval methods}. For clarity, we underline and bold the best performance for each metric. Across both text-to-video and video-to-text retrieval tasks, \sysname consistently outperforms previous SoTA methods on most evaluation metrics.

\begingroup
\renewcommand{\arraystretch}{1.1} 
\setlength{\tabcolsep}{2pt}  
\begin{table*}[htp]
\centering
\belowrulesep=0pt
\aboverulesep=0pt
\resizebox{\textwidth}{!}{ 
\begin{tabular}{c|c|ccc|ccc|ccc|ccc|ccc|ccc|c}
\toprule
\multirow{3}{*}{\textbf{Method}} & \multirow{3}{*}{\textbf{Frames}} & \multicolumn{6}{c|}{\textbf{MSR-VTT}} & \multicolumn{6}{c|}{\textbf{Activity-Net}} & \multicolumn{6}{c|}{\textbf{MSVD}} & \multirow{3}{*}{\makecell[c]{\textbf{FQ-}\\\textbf{Latency(s)}$\downarrow$}}
 \\
\cmidrule{3-20}
& & \multicolumn{3}{c|}{Text $\Rightarrow$ Video} & \multicolumn{3}{c|}{Video $\Rightarrow$ Text} 
& \multicolumn{3}{c|}{Text $\Rightarrow$ Video} & \multicolumn{3}{c|}{Video $\Rightarrow$ Text} 
& \multicolumn{3}{c|}{Text $\Rightarrow$ Video} & \multicolumn{3}{c|}{Video $\Rightarrow$ Text} & \\
& & R@1 & R@5 & MnR$\downarrow$ & R@1 & R@5 & MnR$\downarrow$
& R@1 & R@5 & MnR$\downarrow$ & R@1 & R@5 & MnR$\downarrow$
& R@1 & R@5 & MnR$\downarrow$ & R@1 & R@5 & MnR$\downarrow$ & \\
\midrule
\multicolumn{20}{c}{\textbf{Uniform Sampling Methods}} & \\
\midrule
X-Pool\cite{gorti2022x} & 12/64/12 & 46.7 & 72.8 & 14.3 & 44.4 & 72.6 & 8.5 & 43.6 & 75.8 & 8.0 & 43.2 & 70.6 & 8.1 & 47.2 & 77.4 & \underline{\textbf{9.3}} & 46.3 & 66.9 & 13.4 & 24.18 \\
X-CLIP \cite{ma2022x} & 12/64/12 & 47.0 & 73.6 & 12.1 & 45.3 & 72.7 & 10.1 & 44.3 &74.1 &7.9 &43.9 & 73.9 & 7.6 & 47.1 & \underline{\textbf{77.8}}& 9.5 & \underline{\textbf{60.9}} & \underline{\textbf{87.8}} & \underline{\textbf{4.7}} & 6.77\\
UCoFiA\cite{wang2023unified} & 12/64/12 & \underline{\textbf{49.2}} & 72.0 & 13.0 & 45.6 & 71.0 & 13.4 & \underline{\textbf{45.7}} & 76.0 & 6.6 & 44.6 & 72.8 & 7.5 & \underline{\textbf{47.4}} & 77.6 & 9.6 & 46.5 & 70.4 & 11.2 & 10.12\\
\midrule
\multicolumn{20}{c}{\textbf{Efficient Acceleration Methods}} & \\
\midrule
mmSampler\cite{hu2022mmsampler} & 12/32/12 & 43.4 & 70.9 & 15.9 & 43.3 & 71.2 & 11.0 & 41.7 & 73.0 & 7.4 & 43.4 & 73.8 & 6.9 & 43.0 & 72.5 & 12.6 & 42.1 & 67.4 & 13.1 & 4.46\\
AdaCLIP\cite{hu2023adaclip} & 12/32/12 & 44.6 & 71.7 & 15.5 & 44.0 & 71.9 & 10.7 & 43.5 & 73.7 & 7.4 & 43.7 & 74.5 & 7.1 & 43.5 & 74.2 & 11.2 & 44.6 & 69.1 & 12.7 & 4.55\\
CenterCLIP\cite{zhao2022centerclip} & 12/60/12 & 43.9 & 71.4 & 15.7 & 42.9 & 72.0 & 11.0 & 43.9 & 75.3 & 7.0 & 44.2 & 75.0 & 6.8 & 45.5 & 75.7 & 10.1 & 46.3 & 66.7 & 13.7 & \underline{\textbf{1.89}}\\
EERCF~\cite{tian2024towards} & 12/64/12 & 47.4 & 74.0 & 14.8 & 44.7 & 74.1 & 11.7 & 43.0 & 74.5 & 7.6 & 43.8 & 74.7 & 7.0 & 46.6 & 77.0 & 10.1 & 46.3 & 67.0 & 14.6 & 9.98\\
TEACHCLIP~\cite{tian2024holistic} & 12/64/12 & 46.6 & 74.8 & 13.9 & 44.5 & 72.8 & 11.8 & 42.2 & 72.6 & 8.0 & 43.4 & 74.0 & 7.7 &47.3 & 71.0 &9.8 & 46.9 &67.9 & 12.1 & 6.51\\
\midrule
\textbf{\sysname} (Ours) & 12/32/12 & \textbf{49.0} & \underline{\textbf{75.3}} & \underline{\textbf{11.2}} & \underline{\textbf{47.4}} & \underline{\textbf{76.4}} & \underline{\textbf{7.9}} & \underline{\textbf{45.7}} & \underline{\textbf{76.9}} & \underline{\textbf{6.3}} & \underline{\textbf{46.7}} & \underline{\textbf{78.0}} & \underline{\textbf{5.7}}& \textbf{47.1} & \textbf{77.5} & \textbf{9.5} & \textbf{46.4} & \textbf{67.1} & \textbf{12.6} & \textbf{2.35} \\
\bottomrule
\end{tabular}
}
\caption{Comparison of text-video retrieval methods on MSR-VTT, Activity-Net, and MSVD datasets. Inference efficiency (FQ-Latency) is reported on the MSR-VTT test set.}
\vspace{-0.2cm}
\label{tab:Retrieval}
\end{table*}
\endgroup

On the MSR-VTT dataset, compared to the recent SoTA method UCoFiA\cite{wang2023unified}, \sysname achieves competitive performance on the text-to-video R@1 metric. More impressively, compared to AdaCLIP\cite{hu2023adaclip}, a representative salient frame-based efficient method, our method yields an absolute improvement of nearly 10\% on R@1. Furthermore, in the video-to-text retrieval setting, \sysname outperforms UCoFiA and AdaCLIP by approximately 4\% and 7.7\% on R@1, respectively. \sysname also demonstrates substantial advantages on R@5 and MnR metrics, highlighting its overall retrieval robustness.

On the ActivityNet dataset, which features longer video sequences and more complex paragraph-level queries, \sysname exhibits strong performance. It matches the best result on the text-to-video R@1 metric. Furthermore, it achieves a 4.7\% improvement on video-to-text R@1 over SoTA baselines. Consistent improvements on R@5 and MnR further validate \sysname's effectiveness in aligning long-form visual and textual content. These results suggest that \sysname is well-suited for handling long and semantically rich video content.

On the MSVD dataset, which features short videos with multiple captions per clip, \sysname also achieves competitive results across retrieval tasks, reflecting its generalization ability across varying captioning styles. On the text-to-video task, \sysname achieves comparable R@1 performance to three uniform sampling-based methods, e.g., X-Pool, X-CLIP, and UCoFiA, while significantly reducing model latency, i.e., reducing FQ-inference latency by 90.3\%, 65.3\%, and 76.8\%. Compared to TEACHCLIP, the current SoTA efficient model, \sysname shows a marginal R@1 deficit (47.1 vs. 47.3), but improves R@5 by 9.1\%. More importantly, \sysname reduces FQ-Latency by 63.9\% versus TEACHCLIP, demonstrating superior efficiency for  real-world, scalable edge deployment.

It is worth to note that X-CLIP performs better than \sysname on video-to-text retrieval task. This is mainly because of its exhaustive multi-granular alignment strategy and the loose supervision provided by multiple captions per video. However, video-to-text retrieval is not the primary focus of our work. \sysname is specifically optimized for efficient text-to-video retrieval, where it achieves the same R@1 accuracy and significantly lower inference latency (1.43s, -76\% vs. X-CLIP, see Sec. \ref{sec:efficiency_performance}).


\subsubsection{\textbf{Efficiency Performance}}
\label{sec:efficiency_performance}
To evaluate the practical efficiency of \sysname in real-world scenarios, we conduct an end-to-end retrieval latency test on the MSR-VTT test set, which consists of 1,000 query-video pairs. 

\textbf{Latency comparison.} We conduct an end-to-end retrieval latency comparison between our method and the baselines using a computation ratio of $k{=}50\%$ (See sec \ref{3-3}). As shown in Table~\ref{tab:Retrieval} and Table~\ref{tab:retrieval_performance_msr}, \textbf{\sysname achieves a strong balance between retrieval latency and accuracy}. For FQ-Latency, our method significantly outperforms uniform sampling-based methods such as X-Pool~\cite{gorti2022x}, X-CLIP~\cite{ma2022x}, and UCoFiA~\cite{wang2023unified}. Compared with UCoFiA, \sysname achieves comparable retrieval accuracy in terms of R@1, while reducing FQ-Latency and AQ-Latency by \underline{\textbf{76.8\%}} and \underline{\textbf{75.3\%}}, respectively. Among acceleration-based methods, \sysname also demonstrates superior efficiency. Specifically, it reduces FQ-Latency by \underline{\textbf{48.4\%}} and \underline{\textbf{47.3\%}} compared to salient-frame-based methods mmSampler and AdaCLIP, respectively. Although \sysname’s AQ-Latency is slightly higher than that of CenterCLIP, it achieves significantly better retrieval performance, highlighting a favorable trade-off between speed and effectiveness. 

\textbf{Energy and memory usage.} As shown in Table~\ref{tab:retrieval_performance_msr}, \sysname demonstrates competitive power efficiency, consuming only 475J per average query—comparable to mmSampler (461 J) and significantly lower than most uniform sampling-based methods, such as X-Pool (1603 J), X-CLIP (1103 J), and UCoFiA (2238 J). Although TEACHCLIP achieves the lowest energy usage (235 J), it suffers from lower retrieval accuracy, suggesting a trade-off between power efficiency and retrieval quality.

In terms of memory footprint, \sysname maintains a moderate usage of 5.38GB, which is significantly lower than UCoFiA (12.15 GB) and EERCF (13.85 GB), and remains within practical limits for deployment on mid-end edge GPUs. As our frame encoder also employs MobileNet for fast and efficient visual feature extraction to support relevant frame selection, and its processing differs from that of the CLIP image encoder, the memory for visual processing is about twice as high as X-Pool (2.28 GB) and X-CLIP (1.66 GB) in some cases. Nevertheless, \sysname achieves a favorable balance across latency, energy, and memory dimensions, making it well-suited for edge-aware deployment.

\begingroup
\begin{table}[htp]
\centering
\belowrulesep=0pt
\aboverulesep=0pt
\caption{Retrieval Performance and Efficiency on MSR-VTT Test Set}
\vspace{-5pt}
\renewcommand{\arraystretch}{1.05}
\resizebox{\textwidth}{!}{
\begin{tabular}{c|c|c|c|c|c|c}
\toprule
\textbf{Method} & \textbf{R@1} & \textbf{Frames} & \textbf{FQ-Latency(s)$\downarrow$} & \textbf{AQ-Latency(s)$\downarrow$} & \textbf{AQ Energy(J)$\downarrow$} & \textbf{Max GPU Mem(GB)$\downarrow$} \\
\midrule
X-Pool~\cite{gorti2022x} & 46.7 & 12 & 24.18 & 8.20 & 1603 & 2.28 \\
X-CLIP~\cite{ma2022x} & 47.0 & 12 & 6.77 & 5.92 & 1103 & \underline{\textbf{1.66}} \\
UCoFiA~\cite{wang2023unified} & \underline{\textbf{49.2}} & 12 & 10.12 & 5.80 & 2238 & 12.15 \\
mmSampler~\cite{hu2022mmsampler} & 43.3 & 12 & 4.46 & 0.81 & 461 & 6.49 \\
AdaCLIP~\cite{hu2023adaclip} & 44.6 & 12 & 4.55 & 0.82 & 482 & 4.99 \\
CenterCLIP~\cite{zhao2022centerclip} & 43.9 & 12 & \underline{\textbf{1.89}} & \underline{\textbf{0.38}} & 510 & 4.37 \\
EERCF~\cite{tian2024towards} & 47.6 & 12 & 9.98 & 4.08 & 1741 & 13.85 \\
TEACHCLIP~\cite{tian2024holistic} & 46.7 & 12 & 6.51 & 1.35 & \underline{\textbf{235}} & 4.74 \\
\midrule
\textbf{\sysname} (Ours) & \textbf{49.0} & 12 & \textbf{2.35} & \textbf{1.43} & \textbf{475} & \textbf{5.38} \\
\bottomrule
\end{tabular}
}
\label{tab:retrieval_performance_msr}
\end{table}
\endgroup

\textbf{Computation ratio trade-off.} To determine the optimal computation ratio ($k$\%) for our framework,  we analyze the accuracy-latency trade-off across varying $k$ values, as shown in Table~\ref{tab:retrieval_latency}. Reducing $k$ from from 100\% to 50\% yields 33.4\% FQ-Latency reduction ($3.53s \rightarrow 2.35s$), and 44.6\% AQ-Latency reduction ($2.58s \rightarrow 1.43s$), while preserving competitive retrieval metrics, i.e., R@1 at 49.0 and R@5 at 75.4. This empirically validates \sysname's near-linear speedup without performance degradation, enabled by distillation-driven feature alignment and query-aware semantic-preserving sampling.  

However, aggressive pruning ($k\leq40\%$) induces progressive performance degradation. While R@1 slightly decreases at $k=40\%$ (from 49.0 to 48.7), MnR deteriorates sharply ($13.7 \rightarrow 21.9$), indicating a significant drop in overall ranking quality. This degradation stems from the pruning stage's reduced capacity to retain semantically relevant but low-confidence candidates—a crucial factor for maintaining ranked list integrity. The MnR elevation indicates relevant videos being displaced to lower ranks, particularly impactful in real-world scenarios requiring top-k exploration, where users must navigate extended result lists to locate target content, thereby degrading perceived system responsiveness.

Through comprehensive latency-accuracy pareto analysis,  we select $k=\textbf{50\%}$ as the optimal pruning ratio, which offers the best balance between accuracy and latency. This choice ensures that \sysname delivers a responsive user experience while preserving strong retrieval effectiveness.

\begingroup
\begin{table}[htp]
\centering
\belowrulesep=0pt
\aboverulesep=0pt
\caption{Retrieval performance and latency on the MSR-VTT test set under different computation ratios ($k$\%), where only the top $k$\% of videos are selected for CLIP-based fine-grained computation.}
\vspace{-5pt}
\renewcommand{\arraystretch}{1.0}
\resizebox{0.65\textwidth}{!}{
\begin{tabular}{c|c|c|c|c|c}
\toprule
\textbf{Computation Ratio} & \textbf{R@1}  & \textbf{R@5}  & \textbf{MnR$\downarrow$} & \textbf{FQ-Latency(s)$\downarrow$} & \textbf{AQ-Latency(s)$\downarrow$} \\
\midrule
100\% & 49.0 & 75.3 & 11.2 & 3.53 & 2.58\\
90\%  & 49.0 & 75.4 & 11.2 & 3.33 & 2.37\\
80\%  & 49.0 & 75.4 & 11.2 & 3.08 & 2.13\\
70\%  & 49.0 & 75.4 & 11.6 & 2.80 & 1.87\\
60\%  & 49.0 & 75.4 & 12.5 & 2.58 & 1.64\\
\textbf{50\%} & \textbf{49.0} & \textbf{75.4} & \textbf{13.7} & \textbf{2.35} & \textbf{1.43}\\ \cdashline{1-6}
40\%  & 48.7 & 75.3 & 21.9 & 2.11 & 1.17\\
30\%  & 47.3 & 70.1 & 57.6 & 1.87 & 0.93\\
20\%  & 47.1 & 70.3 & 74.2 & 1.64 & 0.70\\
10\%  & 47.1 & 70.3 & 74.2 & 1.40 & 0.46\\
5\%   & 42.1 & 63.1 & 120.0 & 1.28 & 0.34\\
\bottomrule
\end{tabular}
}
\label{tab:retrieval_latency}
\end{table}
\endgroup

\begin{figure}[hpt]
    \includegraphics[scale=0.3]{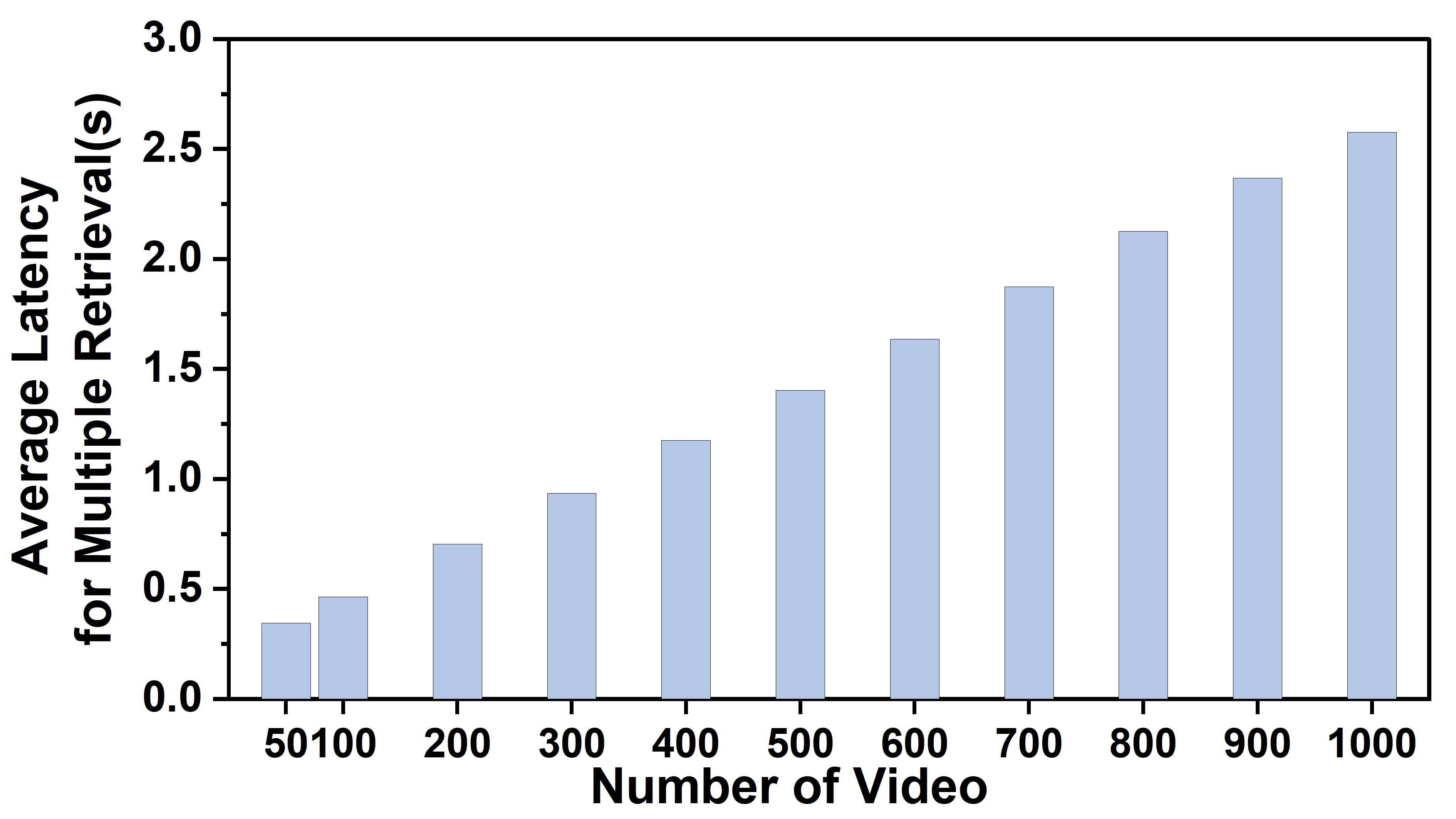}
    \vspace{-1.0em}
    \caption{Retrieval latency under various number of video on the MSR-VTT test set (1,000 videos). }
    \label{video_number_latency}
\end{figure}

\textbf{Analysis of video quantity-retrieval latency.} As shown in Figure \ref{video_number_latency}, the results demonstrate that the retrieval latency of \sysname scales approximately linearly with the number of videos in the dataset. The average latency increases proportionally from 0.34s (50 videos) to 2.58s (1,000 videos), indicating an overall time complexity of $O(n)$ and demonstrating the system’s scalability, efficiency, and predictable computational cost.


\subsection{Ablation Study}
\subsubsection{\textbf{Effectiveness of Prompt-aware Frame Sampling Module}}
We conduct comprehensive ablation studies to evaluate the individual contributions of each component in our prompt-aware sampling strategy. Specifically, we compare four model variants across three benchmark datasets: (1) \textbf{Only-word-level}, which utilizes solely word-level features; (2) \textbf{Only-sentence-level}, which employs only sentence-level features; (3) \textbf{Simple-add}, which sums word-level and sentence-level features; and (4) \textbf{Concat}, which concatenates word-level and sentence-level features. For a fair comparison, all variants include a final linear projection layer following feature integration, ensuring consistent output dimensionality (aligned with our gated fusion design), comparable parameter counts (avoiding under-parameterization in simpler variants), and equivalent model depth in post-fusion processing (i.e., each fusion output is transformed once via a learnable layer).

The ablation results yield two key observations. First, at the feature representation level, word-level features slightly outperform sentence-level ones in text-to-video retrieval, highlighting the importance of fine-grained token-level information for precise visual grounding. Conversely, sentence-level features achieve better performance in video-to-text retrieval, where the task demands alignment with global semantic cues—suggesting the advantage of holistic textual understanding. Second, regarding feature fusion strategies, simple addition surpasses concatenation, potentially due to improved norm preservation during integration. Nevertheless, both static fusion methods underperform compared to our proposed gated fusion, which achieves an improvement of over 3.1 on R@1 metric, emphasizing the value of dynamic, content-aware weighting in multi-granular feature fusion.

\begingroup
\begin{table}[htp]
\centering
\belowrulesep=0pt
\aboverulesep=0pt
\caption{Ablation Study on Prompt-aware Frame Sampling Module.}
\vspace{-5pt}
\renewcommand{\arraystretch}{1.05}
\resizebox{0.7\textwidth}{!}{
\begin{tabular}{c|ccc|ccc|ccc}
\toprule
\multirow{2}{*}{\textbf{Method}} & \multicolumn{3}{c|}{\textbf{MSR-VTT}} & \multicolumn{3}{c|}{\textbf{ActivityNet}} & \multicolumn{3}{c}{\textbf{MSVD}} \\
\cmidrule{2-10}
 & R@1 & R@5 & MnR$\downarrow$ & R@1 & R@5 & MnR$\downarrow$ & R@1 & R@5 & MnR$\downarrow$ \\
\midrule
Word-only & 45.9 & 74.5 & 12.2 & 41.6 & 73.5 & 7.7 & 42.7 & 74.3 & 11.5 \\
Sentence-only & 44.8 & 72.7 & 13.0 & 41.5 & 73.1 & 8.0 & 42.5 & 74.2 & 11.9 \\
Simple-add & 45.9 & 74.1 & 11.5 & 41.7 & 72.8 & 7.7 & 42.8 & 73.9 & 12.1 \\
Concat & 44.8 & 73.0 & 13.1 & 41.9 & 73.3 & 8.0 & 42.5 & 74.0 & 12.0 \\
\midrule
\textbf{Gated (Ours)} & \underline{\textbf{49.0}} & \underline{\textbf{75.3}} & \underline{\textbf{11.1}} & \underline{\textbf{45.7}} & \underline{\textbf{76.9}} & \underline{\textbf{6.3}} & \underline{\textbf{47.0}} & \underline{\textbf{77.5}} & \underline{\textbf{10.3}} \\
\bottomrule
\end{tabular}
}
\label{tab:ablation_study}
\end{table}
\endgroup

\subsubsection{\textbf{Effectiveness of the Distillation Module}}
To assess the impact of the distillation module in the two-stage candidate pruning strategy, we conduct experiments under the 50\% computational budget constraints. The distillation module incorporates a three-layer transformer encoder integrated with the visual feature extractor, as illustrated in the inference stage of Figure~\ref{framework}. For the non-distillation comparison, initial candidate pruning directly computes similarities between textual sentence features and raw visual features from the extractor-temporal modeling pipeline.

As shown in Table~\ref{tab:ablation_distill}, integrating the distillation module enhances performance metrics across all datasets. This improvement substantiates the distillation module's capacity to bridge the representational divergence between lightweight video features and CLIP's cross-modal semantic space. Through feature distribution alignment, the distillation process enables more accurate relevance estimation during coarse pruning, ensuring higher consistency between the distilled candidate set and final CLIP-based ranking.

\begingroup
\renewcommand{\arraystretch}{1.0}
\begin{table}[htp]
\centering
\belowrulesep=0pt
\aboverulesep=0pt
\caption{
Ablation study on Text$\Rightarrow$Video retrieval across three datasets. “w/o Distill.” and “w/ Distill.” denote models without and with the distillation module, respectively. }
\resizebox{0.7\textwidth}{!}{
\begin{tabular}{c|ccc|ccc|ccc}
\toprule
\multirow{2}{*}{\textbf{Method}} 
& \multicolumn{3}{c|}{\textbf{MSR-VTT}} 
& \multicolumn{3}{c|}{\textbf{Activity-Net}} 
& \multicolumn{3}{c}{\textbf{MSVD}} \\
\cmidrule{2-10}
& R@1 & R@5 & MnR$\downarrow$ 
& R@1 & R@5 & MnR$\downarrow$ 
& R@1 & R@5 & MnR$\downarrow$ \\
\midrule
w/o Distill. & 46.6 & 68.9 & 75.9 & 43.9 & 70.2 & 51.8 & 42.6 & 70.8 & 66.7 \\
w/ Distill. & \underline{\textbf{49.0}}& \underline{\textbf{75.4}} & \underline{\textbf{13.7}} & \underline{\textbf{47.4}}& \underline{\textbf{76.2}}& \underline{\textbf{8.2}} & \underline{\textbf{47.0}}& \underline{\textbf{77.5}} & \underline{\textbf{10.3}} \\
\bottomrule
\end{tabular}
}
\label{tab:ablation_distill}
\end{table}
\endgroup

It is worth noting that the performance degradation observed under the 50\% computation ratio in the non-distillation setting stems from the fact that the visual features used for initial pruning are not explicitly trained to align with CLIP's semantic space. These features are only involved in optimizing the prompt-aware frame sampling module, which focuses on identifying relevant frames rather than learning CLIP-aligned representations. As a result, the candidate pruning based on such features becomes less reliable, leading to lower recall.

\subsection{\textbf{Personalized Retrieval Performance Evaluation}} 
\label{sec:personalized evaluation}
\subsubsection{\textbf{Personalized Query Evaluation on ActivityNet.}}
To validate our prompt-aware sampling strategy's capability in handling real-world query diversity, we test its performance on a personalized subset of the ActivityNet dataset. Specifically, we randomly sample 100 videos from ActivityNet and invite 5 volunteers to manually create personalized text queries for each video, reflecting a wide range of individual interests and perspectives. Unlike the canonical queries provided in the dataset, these new queries not only focus on different parts of the video content but also include fine-grained subjective aspects that align with each user's personal preferences. This protocol systematically captures the multi-dimensional complexity of authentic user information needs.

\begingroup
\renewcommand{\arraystretch}{1.0}
\begin{table}[htp]
\centering
\caption{Performance comparison of text-video and video-text retrieval on a personalized subset of the ActivityNet.}
\vspace{-2pt}
\belowrulesep=0pt
\aboverulesep=0pt
\resizebox{0.7\textwidth}{!}{
\begin{tabular}{c|c|ccc|ccc}
\toprule
\multirow{2}{*}{\textbf{Method}} & \multirow{2}{*}{\textbf{Frames}} & \multicolumn{3}{c|}{Text $\Rightarrow$ Video} & \multicolumn{3}{c}{Video $\Rightarrow$ Text} \\
& & R@1 & R@5 & MnR$\downarrow$ & R@1 & R@5 & MnR$\downarrow$ \\
\midrule
X-Pool\cite{gorti2022x} & 64 & 97.9 & 100.0 & 1.5 & 90.4 & 100.0 & 2.3 \\
X-CLIP \cite{ma2022x} & 64 & 98.0 & 100.0 & 1.1 & 89.7 & 98.5 & 5.5 \\
UCoFiA\cite{wang2023unified} & 64 & 100.0 & 100.0 & 1.0 & \underline1{\textbf{91.2}} & 100.0 & \underline{\textbf{1.0}} \\
mmSampler\cite{hu2022mmsampler} & 32 & 60.8 & 86.2 & 15.5 & 61.3 & 88.0 & 9.8 \\
AdaCLIP\cite{hu2023adaclip} & 32 & 62.7 & 87.0 & 11.2 & 61.5 & 89.4 & 8.8 \\
CenterCLIP\cite{zhao2022centerclip} & 60 & 61.5 & 85.5 & 12.0 & 63.6 & 86.4 & 9.0 \\
EERCF\cite{tian2024towards} & 64 & 83.3 & 100.0 & 1.3 & 86.7 & 100.0 & 1.3 \\
TEACHCLIP\cite{tian2024holistic} & 64 & 89.3 & 100.0 & 1.5 & 90.7& 100.0 & 1.2 \\
\midrule
\textbf{\sysname} (Ours) & 32 & \underline{\textbf{100.0}} & \underline{\textbf{100.0}} & \underline{\textbf{1.0}} & \textbf{90.7} & \underline{\textbf{100.0}} & \textbf{1.1} \\
\bottomrule
\end{tabular}
}
\label{tab:30activitynet_results}
\end{table}
\endgroup

As evidenced in Table~\ref{tab:30activitynet_results}, \sysname achieves strong performance in both text-to-video and video-to-text retrieval tasks, with high R@1 scores and low MnR, demonstrating its adaptability to diverse user queries. Analysis reveals that baseline models employing dense frame sampling, e.g., X-CLIP, X-Pool, and UCoFiA, also attain competitive accuracy, primarily due to two factors: (1) dense frame sampling (64 frames) ensures comprehensive temporal coverage, and (2) text-aware fusion or alignment strategies enhance associations between visual content and user queries. However, these approaches incur substantial computational overhead.

In contrast, mmSampler, AdaCLIP, and CenterCLIP  exhibit significantly degraded performance, with R@1 scores dropping to 60.8, 62.7, and 61.5 respectively, alongside increased MnR. This performance decline stems from their reliance on static sampling criteria that cannot adapt to user-specific queries. Since the selected static salient frames often exclude content relevant to user interests, the resulting video features misalign with query semantics, leading to low similarity scores and poor retrieval outcomes.

\subsubsection{\textbf{User Satisfaction Evaluation}} 
To systematically assess the practical utility of text-video retrieval methods, we conducted a user study involving 100 participants from diverse age groups, professions, and genders. The study spanned two weeks and was carried out on participants’ personal computers, equipped with GPUs ranging from RTX 2080Ti to RTX 5070, reflecting realistic deployment environments.

The user study comprised two main tasks: (1) Public Benchmark Retrieval: Participants browsed the MSR-VTT and ActivityNet datasets and constructed 20 query-video pairs each. These were then used to evaluate retrieval performance across all baseline methods, ensuring consistent and fair comparison. (2) Personal Video Library Retrieval: Participants utilized their own video collections—including self-recorded content and videos downloaded from platforms such as YouTube or TikTok (each under 3 minutes, with 200–500 videos per participant)—to create another 20 query-video pairs. This task assessed retrieval quality in real-world, personalized scenarios.

\begin{figure}[hpt]
    \includegraphics[scale=0.62]{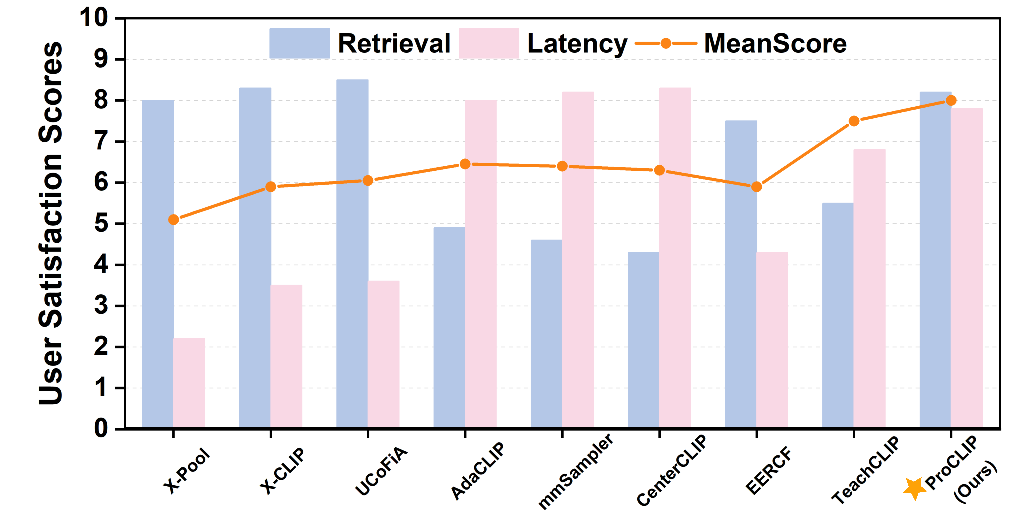}
    \vspace{-1.0em}
    \caption{User satisfaction scores.}
    \label{user study}
\end{figure}

To quantify user perceptions, each participant rated their retrieval experience for every query using a 10-point scale (10 indicating optimal experience, 0 indicating complete failure), jointly evaluating retrieval accuracy and system responsiveness. As shown in Figure \ref{user study}, dense frame sampling methods (e.g., X-Pool, X-CLIP and UCoFiA) achieved relatively high ratings in retrieval accuracy but suffered severe system latency, significantly degrading user experience. Rapid retrieval methods (e.g., AdaCLIP, mmSampler, CenterCLIP and TEACHCLIP) delivered sub-second response latency but exhibited frequent failure in critical semantic matching, resulting in markedly limited overall satisfaction. In contrast, \sysname achieved optimal balance between retrieval accuracy and response efficiency. Its comprehensive satisfaction scores significantly outperformed all baseline methods, validating its technical advantages in real-world deployment scenarios.

\section{Conclusion}


In this paper, we propose \sysname, a user-centric and edge-friendly framework for efficient text-video retrieval. \sysname introduces a prompt-aware frame sampling strategy that dynamically selects semantically relevant frames conditioned on user queries, addressing the query-insensitive limitations of conventional static sampling approaches. Furthermore, \sysname employs a two-stage candidate pruning mechanism to significantly reduce computational overhead while preserving retrieval accuracy. Extensive experiments on three public benchmarks and two personalized evaluation settings demonstrate that \sysname strikes a strong balance between retrieval performance and latency, making it well-suited for real-world applications on edge devices. Moving forward, we plan to develop an adaptive algorithm to automatically determine the optimal computation ratio $k\%$ based on input characteristics, further enhancing the usability, adaptability, and robustness of our framework in practical deployment scenarios.

\bibliographystyle{ACM-Reference-Format}
\bibliography{reference}

\end{document}